\documentclass[twocolumn,showpacs,aps,prl,amsmath,amssymb]{revtex4}

\usepackage{graphicx}
\usepackage{dcolumn}
\usepackage{bm}
\usepackage{amsmath,epsfig}

\newcommand{\ket}[1]{\left\vert#1\right\rangle}
\newcommand{\s}{\uparrow}
\newcommand{\g}{\downarrow}

\begin{document}

\author{Francesco Ciccarello\mbox{$^{1,2}$}}
\email {ciccarello@difter.unipa.it}

\author{G. Massimo Palma \mbox{$^{2}$}, Michelangelo Zarcone \mbox{$^{1}$},
Yasser Omar\mbox{$^{3}$}, Vitor Rocha Vieira\mbox{$^{4}$}}
\affiliation{ \mbox{$^{1}$} CNISM and Dipartimento di Fisica e
Tecnologie Relative dell'Universit\`{a} degli Studi di Palermo,
Viale delle Scienze,
Edificio 18, I-90128 Palermo, Italy \\
\mbox{${\ }^{2}$}NEST- INFM (CNR) \& Dipartimento di Scienze Fisiche
ed Astronomiche dell'Universit\`{a}
degli Studi di Palermo, Via Archirafi 36, I-90123 Palermo, Italy \\
\mbox{${\ }^{3}$} SQIG, Instituto de Telecomunica\c{c}\~oes,
P-1049-001 Lisbon and CEMAPRE, ISEG, Technical University of Lisbon, P-1200-781 Lisbon, Portugal  \\
\mbox{${\ }^{4}$}Department of Physics, Instituto Superior
T\'{e}cnico, Av. Rovisco Pais, 1049-001 Lisbon, Portugal}

\begin{abstract}
In a recent paper -- F. Ciccarello \emph{et al.}, New J. Phys.
\textbf{8}, 214 (2006) -- we have demonstrated that the electron
transmission properties of a one-dimensional (1D) wire with two
identical embedded spin-1/2 impurities can be significantly affected
by entanglement between the spins of the scattering centers. Such
effect is of particular interest in the control of transmission of
quantum information in nanostructures and can be used as a detection
scheme of maximally entangled states of two localized spins. In this
letter, we relax the constraint that the two magnetic impurities are
equal and investigate how the main results presented in the above
paper are affected by a static disorder in the exchange coupling
constants of the impurities. Good robustness against deviation from
impurity symmetry is found for both the entanglement dependent
transmission and the maximally entangled states generation scheme.
\end{abstract}



\title{Effect of Static Disorder in an Electron Fabry-Perot Interferometer with Two Quantum Scattering Centers}

\maketitle

The key role that entanglement plays in quantum information
processing has been investigated over the past few years
\cite{nielsen}. In this framework, the role that it plays in quantum
transport in mesoscopic systems has been analyzed
\cite{quantum_transp}. Recently, we have shown a novel way in which
entanglement can be used for controlling electron transport in
nanostructures \cite{ciccarello}. Assume to have a 1D wire where two
spin-1/2 impurities are embedded at a fixed distance. Such system
can be regarded as the electron analogue of a Fabry-Perot (FP)
interferometer, with the impurities playing the role of two mirrors
with a spin quantum degree of freedom. Single electrons are injected
into the wire and undergo multiple scattering between the two
magnetic impurities due to the presence of a contact exchange
electron-impurity coupling. At each scattering event spin-flip may
occur and thus the transmitted spin state of the overall system will
be generally different from the incoming one. The typical behaviour
shown by electron transmittivity $T$ consists of a loss of electron
coherence and thus of a resonance condition $T=1$, due to the
presence of internal spin degrees of freedom of the scattering
centers \cite{imry_joshi}. Such a system is indeed the electron
analogue of a Fabry-Perot (FP) interferometer, with the impurities
playing the role of two mirrors with a spin quantum degree of
freedom. However, unlike the standard FP device where scattering
with each mirror introduces a well-fixed phase shift, in the present
system the above phase shifts depend on the electron-impurities spin
state and thus, in general, a resonance condition cannot take place.
However, the presence of quantum scatterers allows one to
investigate if and to what extent maximally entangled states of the
impurity spins can affect electron transmission. Denoting by
$\ket{\Psi^{\pm}}=2^{-\frac{1}{2}}(\ket{\s\g}\pm\ket{\g\s})$ the
triplet and singlet maximally entangled spin states of the
impurities, respectively, we have thus found that when
$\ket{\Psi^-}$ is prepared, a perfect resonance condition $T=1$ can
be always reached at electron wave vectors fulfilling $kx_0=n\pi$
($n$ integer, $x_0$ the distance between the impurities) and
regardless of the electron spin state. When this occurs the incoming
spin state of the electron-impurities system is transmitted
completely unchanged. Therefore, a sort of perfect ``transparency"
takes place \cite{ciccarello}. Moreover, as illustrated in Fig. 1,
electron transmission within the one spin-up family of impurity
states $\cos\vartheta\ket{\s\g} +
e^{i\varphi}\sin\vartheta\ket{\g\s}$ is maximized (minimized) by
$\ket{\Psi^-}$ ($\ket{\Psi^+}$). $T$ is thus crucially affected by
the relative phase $\varphi$.
\begin{figure}
 \includegraphics [scale=0.9]{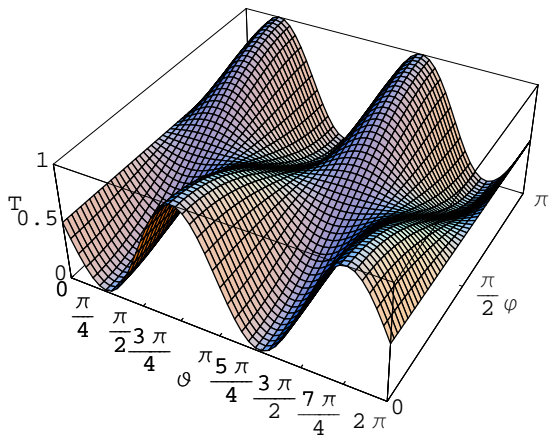}
 \caption{\label{teta_phi} Electron transmittivity $T$ at $kx_{0}=n\pi$ and
$\rho(E)J=10$
 when the electron is injected in an arbitrary spin
state with the impurities prepared in the state
$\cos{\vartheta}\ket{\s\g}+e^{i\varphi}\sin{\vartheta}\ket{\g\s}$.
$\rho(E)$ is the density of states per unit length of the wire.}
\label{fig4}
\end{figure}
This suggests the appealing possibility to use entanglement between
the impurity spins for controlling electron transmission in a 1D
wire or, alternatively, for implementing a maximally entangled
states detection scheme via electron transmission. The above
phenomena have been demonstrated to follow from an effective
conservation law occurring whenever the condition $kx_0=n\pi$ is
fulfilled. In \cite{ciccarello} the following Hamiltonian has been
assumed
\begin{equation} \label{H}
H=\frac{p^{2}}{2m^*}+ J \, \mbox{\boldmath$\sigma$}\cdot
\mathbf{S}_{1} \,\delta(x)+ J \, \mbox{\boldmath$\sigma$} \cdot
\mathbf{S}_{2}\,\delta(x-x_{0})
\end{equation}
where $p=-i \hbar \nabla$, $m^*$ and $\bm{\sigma}$  are the electron
momentum operator, effective mass and spin-1/2 operator
respectively, $\mathbf{S}_{i}$ ($i=1,2$) is the spin-1/2 operator of
the $i$-th impurity and  $J$ is the exchange spin-spin coupling
constant between the electron and each impurity. Denoting by
$\mathbf{S}=\bm{\sigma}+\mathbf{S}_{1}+\mathbf{S}_{2}$ and
$\mathbf{S}_{12}=\mathbf{S}_{1}+\mathbf{S}_{2}$ the total spin of
the system and of the two impurities, respectively,  Hamiltonian
(\ref{H}) implies conservation of $\mathbf{S}^2$ and $S_z$.
$\mathbf{S}_{12}^2$ is in general not conserved due to the
difference between $\delta(x)$ and $\delta(x-x_0)$ in (\ref{H}).
However, when $kx_0=n\pi$ the effective representations
$\delta_{k}(x)$ and $\delta_{k}(x-x_0)$ of these two electron
orbital operators coincide (the electron being found at $x=0$ and
$x=x_0$ with equal probability) and $\mathbf{S}_{12}^2$ turns out to
be an additional constant of motion. This fact is the ultimate
reason for the occurrence of the above mentioned behaviours
associated with $\ket{\Psi^-}$ and $\ket{\Psi^+}$ (note that these
are eigenstates of $\mathbf{S}_{12}^2$), as explained in detail in
\cite{ciccarello}.

However, the above effective conservation law of $\mathbf{S}_{12}^2$
relies on the assumption of dealing with two perfectly identical
impurities with equal coupling constant $J$. Of course, due to
unavoidable static disorder, this condition cannot be strictly
realized in a real system. The aim of this paper is to investigate
how large the difference between the coupling constants of the two
impurities can be before the entanglement dependent effects and the
maximally entangled states generation scheme presented in
\cite{ciccarello} are significantly spoiled.

To begin with, let $J_i$ ($i=1,2$) be the exchange coupling constant
of the $i$-th impurity. We thus generalize Hamiltonian (\ref{H}) as
\begin{equation} \label{H_j1_j2}
H=\frac{p^{2}}{2m^*}+ J_1 \, \mbox{\boldmath$\sigma$}\cdot
\mathbf{S}_{1} \,\delta(x)+ J_2 \, \mbox{\boldmath$\sigma$} \cdot
\mathbf{S}_{2}\,\delta(x-x_{0})
\end{equation}
It is convenient to introduce the quantities $\bar{J}=(J_1+J_2)/2$
and $\Delta J=J_2-J_1$ through which $J_i$ ($i=1,2$) can be
expressed as $J_1=\bar{J}-\Delta J/2$ and $J_2=\bar{J}+\Delta J/2$.
Our previous results with identical impurities are thus recovered
for $\Delta J \rightarrow 0$. The exact stationary states of the
system at all orders in $J_1$ and $J_2$ can be derived through an
appropriate quantum waveguide theory approach. Since $\mathbf{S}^2$
and $S_z$ are still constants of motion when $J_1\neq J_2$, the
block diagonalization-based procedure used for the case of two
identical impurities \cite{ciccarello} can be readopted, the
difference being that in the present case there is an additional
parameter. Denoting the total spin of the electron and the $i$-th
impurity as $\mathbf{S}_{ei}=\bm{\sigma}+\mathbf{S}_{i}$ and
assuming left-incident electrons, we use as spin-space basis the
states $\ket{s_{e2}; s, m_s}$, common eigenstates of
$\mathbf{S}_{e2}^{2}$, $\mathbf{S}^{2}$ and $S_z$, to express, for a
fixed wave vector $k>0$, each of the eight stationary states of the
system as an 8D column. The calculation of the stationary states
through suitable boundary conditions \cite{ciccarello} allows us to
find all the transmission probability amplitudes
$t_{s_{e2}}^{(s'_{e2},s)}$ that an electron prepared in the incoming
state $\ket{k}\ket{s'_{e2}; s, m_s}$ is transmitted in a state
$\ket{k}\ket{s_{e2}; s, m_s}$. These coefficients can be used to
compute how an electron is transmitted through the wire for any
arbitrary initial spin state of the system \cite{ciccarello}. The
transmission amplitudes $t_{s_{e2}}^{(s'_{e2},s)}$ turn out to be
functions of the three dimensionless parameters $kx_0$,
$\rho(E)\bar{J}$ and $\rho(E)\Delta J$, where
$\rho(E)=(\sqrt{2m^{*}/E})/\pi\hbar$ is the density of states per
unit length of the wire as a function of the electron energy $E$.

We begin our analysis by investigating how the effect of perfect
transparency shown when the impurity spins are prepared in the
singlet state is affected by a difference in the two coupling
constants. In Fig. 2 we plot the electron transmittivity $T$ versus
$\rho(E)\bar{J}$ and $\rho(E)\Delta J$ when the electron is injected
in an arbitrary spin state for $kx_0=n\pi$ with the impurities in
the state $\ket{\Psi^-}$.
\begin{figure}
 \includegraphics [scale=0.9]{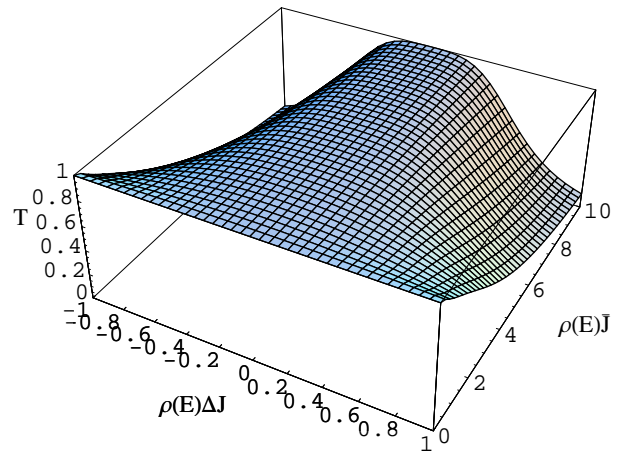}
 \caption{\label{teta_phi} $T$ versus $\rho(E)\bar{J}$ and $\rho(E)\Delta J$ for
$kx_0=n\pi$ when the electron is injected in an arbitrary spin state
with the impurities prepared in the state $\ket{\Psi^-}$.
$\rho(E)\Delta J$ is normalized to $\rho(E)\bar{J}$, reducing to the
ratio $\Delta J/\bar{J}$.}
\end{figure}
Note that $T=1$ for $\Delta J=0$ since the case of perfect
transparency with identical impurities is recovered. As expected,
for a fixed $\rho(E)\bar{J}$, $T$ decreases for increasing values of
$|\Delta J|$ due to the progressive lack of conservation of
$\mathbf{S}_{12}^2$. Note how this decrease gets faster for
increasing strengths of $\rho(E)\bar{J}$, indicating that for a
given electron energy low coupling constants $\bar{J}$ show better
robustness against impurities'asymmetry. It turns out that for a
difference in the coupling constants larger than 25\% compared to
$\bar{J}$, perfect transparency is not significantly spoiled
($T>0.95$) in the whole broad range of strengths of $\rho(E)\bar{J}$
here considered.

A remarkable feature of the plot in Fig. 2 is its symmetry with
respect to a change of sign in $\Delta J$ for a fixed $\bar{J}$ (the
relevant parameter being thus $|\Delta J|$). To explain this, we
recall that, for $kx_0=n\pi$, $\delta_{k}(x)=\delta_{k}(x-x_0)$ and
write the non-kinetic part $V$ of Hamiltonian (\ref{H_j1_j2}) in the
form
\begin{equation} \label{H_j1_j2_2}
V=\left[ \bar{J} \, \mbox{\boldmath$\sigma$}\cdot
(\mathbf{S}_{1}+\mathbf{S}_{2})+\frac{\Delta J}{2}
\mbox{\boldmath$\sigma$}\cdot
(\mathbf{S}_{2}-\mathbf{S}_{1})\right]\,\delta_{k}(x)
\end{equation}
Note that in such regime, where the electron has equal probability
to be found at $x=0$ and at $x=x_0$, a change in the sign of $\Delta
J$ is equivalent to an interchange of the impurity indexes.
Therefore, the above symmetry property of $T$ in the case of Fig. 2
straightforwardly follows from the symmetry of $\ket{\Psi^-}$ under
an interchange of impurity 1 and 2. In the remaining of this paper,
we will thus consider only positive values of $\Delta J$ whenever
the initial spin state is symmetric for an interchange of the two
impurities, such in the cases of $\ket{\Psi^+}$ and $\ket{\g\g}$.

Denoting by $T_{\Psi^{\pm}}$ the value of $T$ obtained for
$\ket{\Psi^{\pm}}$, it is worth analyzing the behaviour of $\Delta
T=T_{\Psi^{-}}-T_{\Psi^{+}}$, that is the difference of the electron
transmittivity for $\ket{\Psi^-}$ (high transmission) and
$\ket{\Psi^+}$ (low transmission). In order to observe the
entanglement controlled-transmittivity and/or detect maximally
entangled singlet/triplet states, one aims at having the highest
possible value of $\Delta T$ \cite{ciccarello} with the hope that it
is only weakly affected by some impurities' asymmetry. Regarding the
latter issue, in Fig. 3 we plot $\Delta T$ normalized to its value
for $\Delta J=0$ versus $\rho(E)\bar{J}$ and $\rho(E)\Delta J$ in
the same regime considered in Fig. 2.
\begin{figure}
 \includegraphics [scale=0.9]{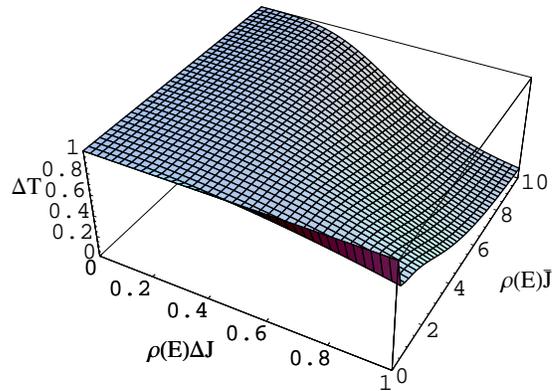}
 \caption{\label{teta_phi} $\Delta T=T_{\Psi^{-}}-T_{\Psi^{+}}$ versus
$\rho(E)\bar{J}$ and $\rho(E)\Delta J$ for $kx_0=n\pi$ when the
electron is injected in an arbitrary spin state. $\rho(E)\Delta J$
is normalized to $\rho(E)\bar{J}$, reducing to the ratio $\Delta
J/\bar{J}$. $\Delta T$ is normalized to its value for $\Delta J=0$.}
\end{figure}
A behaviour qualitatively very similar to the case of Fig. 2 is
exhibited. In the whole considered range of $\rho(E)\bar{J}$,
$\Delta T$ turns out to be only weakly affected ($\Delta T>0.95$) by
a difference in the impurity coupling constants up to more than
25\%.

To observe the entanglement-dependent electron transmittivity one
must of course be able to prepare the maximally entangled states
$\ket{\Psi^-}$ and $\ket{\Psi^+}$. These states can be easily
transformed into each other by simply introducing a relative phase
shift through a local field. In \cite{ciccarello} we have proposed a
scheme to generate the state $\ket{\Psi^+}$ via electron scattering,
improving a previous recent proposal \cite{yasser}. The idea is to
inject an electron in the state $\ket{\s}$ in the regime $kx_0=n\pi$
with the two impurity spins initially in the state $\ket{\g\g}$. Due
to conservation of both $\mathbf{S}_{12}^2$ and $S_z$, when the
electron is transmitted in the state $\ket{\g}$ with probability
$T_{\g}$, the two impurities are projected into the state
$\ket{\Psi^{+}}$ \cite{ciccarello}. A difference in the coupling
constants of the impurities is expected to modify the spin-polarized
transmission probability $T_{\g}$. Moreover, since the scheme relies
on the conservation of $\mathbf{S}_{12}^2$, the two localized spins
in general will not be projected in the maximally entangled state
$\ket{\Psi^{+}}$. In Figs. 4 and 5 we thus plot respectively
$T_{\g}$ and the fidelity $F_{\Psi^+}$ with respect to
$\ket{\Psi^+}$ of the (normalized) spin state into which the
impurities are projected after the electron is transmitted in the
spin down state.
\begin{figure}
 \includegraphics [scale=0.9]{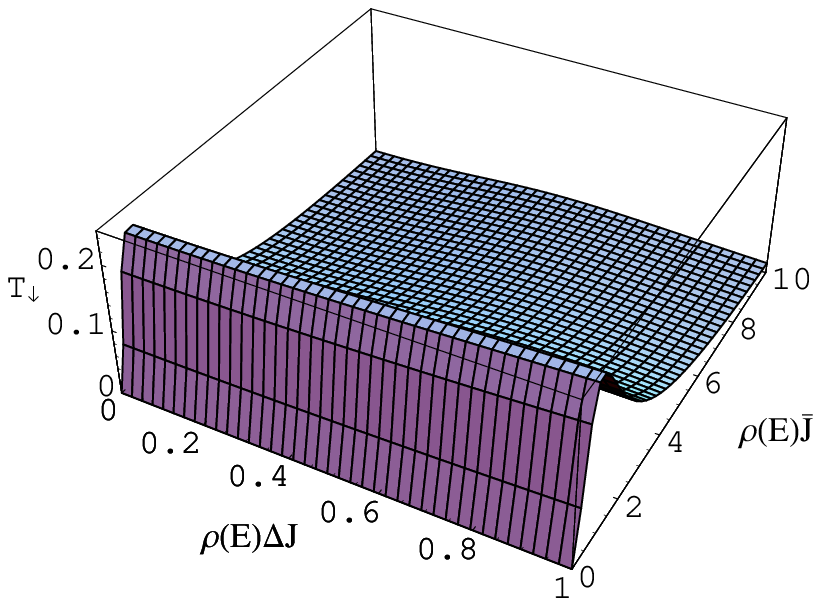}
 \caption{$T_{\g}$ versus $\rho(E)\bar{J}$ and $\rho(E)\Delta J$ for
$kx_0=n\pi$ when the electron is injected in the state $\ket{\s}$
with the impurities prepared in the state $\ket{\g\g}$.
$\rho(E)\Delta J$ is normalized to $\rho(E)\bar{J}$, reducing to the
ratio $\Delta J/\bar{J}$.}
\end{figure}
\begin{figure}
 \includegraphics [scale=0.9]{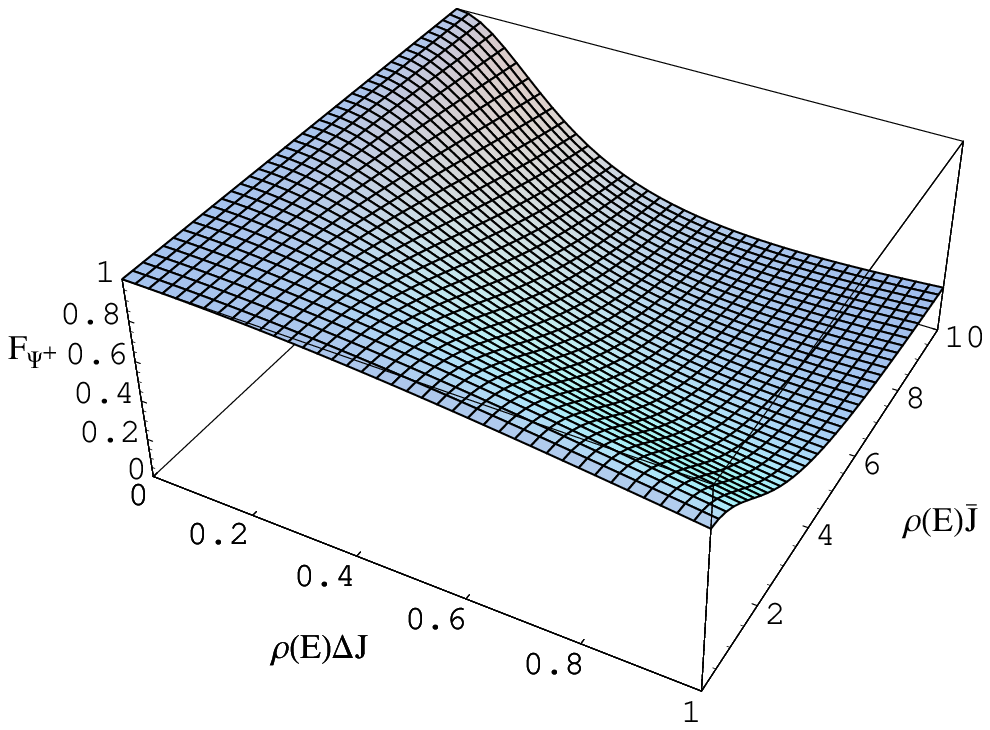}
 \caption{$F_{\Psi^+}$ versus $\rho(E)\bar{J}$ and $\rho(E)\Delta J$
for $kx_0=n\pi$ when the electron is injected in the state
$\ket{\s}$ with the impurities prepared in the state $\ket{\g\g}$.
$\rho(E)\Delta J$ is normalized to $\rho(E)\bar{J}$, reducing to the
ratio $\Delta J/\bar{J}$.}
\end{figure}
$T_{\g}$ is almost negligibly affected by the presence of $\Delta
J$. The same is not true for $F_{\Psi^+}$ which is indeed expected
to be very sensible to the lack of conservation of
$\mathbf{S}_{12}^2$. $F_{\Psi^+}>0.95$ in the whole considered range
of $\rho(E)\bar{J}$ for a difference in the impurity coupling
constants up to more than 5\%. However, for $\rho(E)\bar{J}\simeq
1$, that is the strength of the impurity coupling constant
maximizing $T_{\g}$, $F_{\Psi^+}>0.95$ up to values of $|\Delta
J|/\bar{J}$ larger than 30\%.

In conclusion, the results presented in this paper show very good
tolerance of the entanglement-dependent transmission effects
occurring in a 1D wire with two spin-1/2 impurities
\cite{ciccarello} with respect to unavoidable static disorder in the
coupling constants of the impurities. Therefore, the experimental
difficulty in realizing two perfectly identical magnetic impurities
does not appear to be an obstacle for the observation of such
phenomena.

\textbf{Acknowledgements} Helpful discussions with J.-M. Lourtioz
and G. Fishman are gratefully acknowledged. The authors thank the
support from CNR (Italy) and GRICES (Portugal). YO and VRV thank the
support from Funda\c{c}\~{a}o para a Ci\^{e}ncia e a Tecnologia
(Portugal), namely through programs POCTI/POCI and projects
POCI/MAT/55796/2004 QuantLog and POCTI-SFA-2-91, partially funded by
FEDER (EU).


\begin{thebibliography}{10}
\bibitem{nielsen} M. A. Nielsen, and I. L. Chuang, \textit{Quantum Computation and
Quantum Information}, Cambridge Univ. Press, Cambridge, (2000)
\bibitem{quantum_transp}  G. Burkhard , D. Loss, and E. V. Sukhorukov,  Phys. Rev. B
\textbf{61}, R16303 (2000); P. Samuelsson, E. V. Sukhorukov and M.
B\"{u}ttiker, Phys. Rev. B \textbf{70}, 115330 (2004); F. Taddei and
R. Fazio, Phys. Rev. B \textbf{65}, 075317 (2002).
\bibitem{ciccarello}  F. Ciccarello, G. M. Palma, M. Zarcone, Y. Omar and V. R. Vieira, New J. Phys.
\textbf{8}, 214 (2006); F. Ciccarello, G. M. Palma, M. Zarcone, Y.
Omar and V. R. Vieira, J. Phys. A: Math. Theor. \textbf{40}, 7993
(2007)
\bibitem{imry_joshi}  A. Stern, Y. Aharonov, and Y. Imry, Phys. Rev. A \textbf{41}, 3436
(1990); S. K. Joshi, D. Sahoo, and A. M. Jayannavar, Phys. Rev. B
\textbf{64}, 075320 (2001).
\bibitem{yasser} A. T. Costa, Jr., S. Bose, and Y. Omar, Phys. Rev. Lett. \textbf{96},
230501 (2006).
\end{thebibliography}
\end{document}